\documentclass[prl,twocolumn,superscriptaddress,showpacs]{revtex4}
\usepackage{graphicx}
\usepackage{epsfig}

\usepackage{amssymb,amsmath,amsfonts,hyperref}
\usepackage{wasysym}
\usepackage{latexsym}

\newcommand{\nn}{\nonumber}

\newcommand{\be}{\begin{eqnarray}}
\newcommand{\ee}{\end{eqnarray}}

\begin{document}
\preprint{APS/123-QED}
\title{Magnetization plateaus and sublattice ordering in easy axis Kagome lattice antiferromagnets}
\author{Arnab Sen}
\address{{\small Department of Theoretical Physics,
Tata Institute of Fundamental Research,
Homi Bhabha Road, Mumbai 400005, India}}
\author{Kedar Damle}
\address{{\small Department of Theoretical Physics,
Tata Institute of Fundamental Research,
Homi Bhabha Road, Mumbai 400005, India}}
\address{{\small Physics Department, Indian Institute of Technology
Bombay, Mumbai 400076, India}}
\author{Ashvin Vishwanath}
\address{{\small Department of Physics, University of California, Berkeley, CA 74720}}

\date{\today}
\begin{abstract}
We study kagome lattice antiferromagnets where the effects of easy
axis single-ion anisotropy ($D$) dominates over the Heisenberg
exchange $J$. For $S \ge 3/2$, virtual quantum fluctuations help
lift the extensive classical degeneracy. We demonstrate the presence of a one-third magnetization plateau
for a broad range of
magnetic fields $J^3/D^2 \lesssim B \lesssim JS$ along the easy axis. The fully
equilibriated system at low temperature on this plateau develops an unusual {\em nematic} order that breaks sublattice
rotation symmetry but not translation symmetry---however, extremely slow dynamics associated
with this ordering is expected to lead to glassy freezing of the system on intermediate time-scales.

\end{abstract}

\pacs{75.10.Jm,75.10.Dg}
\vskip2pc

\maketitle

 Geometrically frustrated  magnets, which are characterized by a
large number of symmetry unrelated {\em classical} ground states,
display a wealth of new phenomena. In contrast to their
unfrustrated counterparts, in which the low temperature ordering is largely
determined by classical energetics even for small spin length $S$, quantum effects can play a
crucial role in frustrated magnets, precisely because the classical
energetics fails to pick a unique state.
Of particular interest are those systems
in which a broad cooperative paramagnetic
regime~\cite{Moessnerreview} at intermediate temperatures gives way
at low temperature to a variety of novel ordered and liquid
phases~\cite{Misguich} arising from the quantum fluctuations of
spins. Even when magnetic order is selected, the resulting patterns
are often complex, and are promising candidates for realizing
multiferroic properties \cite{Mostovoy}.

The Kagome lattice composed of corner sharing triangles is one of
the the most frustrated lattice arrangements possible, and one that occurs commonly in nature.
Ground states of the
antiferromagnetic Heisenberg model on this lattice typically involve
coplanar arrangement of spins, and are extensively degenerate in the
limit of classical spins~\cite{Rutenberg_Huse}. Despite much work on
the role of quantum fluctuations in lifting this degeneracy, several
fundamental questions regarding the ground state of low spin Kagome
antiferromagnets (eg. $S=1/2,1,3/2 \dots$)  remain to be decisively
settled~\cite{Misguich}.

In some cases however, {\em collinear} spin arrangements are
preferred, for example in the presence of strong single-ion
anisotropy that leads to an easy axis. This occurs in the recently
studied Nd$_3$Ga$_5$SiO$_{14}$ (NGS) compound
~\cite{kagome1,kagome2}, where Nd$^{3+}$ (total angular momentum
${\bf J}=9/2$) ions form a kagome antiferromagnet in which the
common $c$ axis is the easy axis at low temperature (below $33K$).
The classical ground states of such collinear spins are also
extensively degenerate. In contrast to the isotropic case, the
effect of quantum fluctuations in selecting the low temperature state can be
studied in a much more controlled fashion, and this is the focus of our work here.

Our
results are readily stated. First, the selection mechanism depends
strongly on the spin length $S$. Below a critical spin length
$S_c=3/2$, quantum transitions between different classical ground
states dominate the degeneracy splitting. This `kinetic energy'
dominated selection applies only to $S=1$ ($S=1/2$ does not admit a
single ion anisotropy): 
There, a uniform spin nematic ground state was found in zero field,
and a one-third magnetization ($m=1/3$) plateau with
$\sqrt{3}\times\sqrt{3}$ collinear order obtains in the presence of
a magnetic field directed along the easy axis~\cite{Damle_Senthil}.

As we show here, the situation for all $S\geq S_c=3/2$ is entirely
different. In these cases, virtual
quantum transitions dominate the energetics, leading to `potential
energy' differences between different classical ground states.
The $m=1/3$ plateau, which is found to exist over a broad range of
fields, displays an unusual {\em nematic} order at low
temperature, {\em i.e.}, it spontaneously breaks sublattice rotation symmetry
but not translation symmetry (or spin rotation symmetry about the
field direction). Furthermore, the system exhibits slow {\em glassy}
dynamics in this state as a consequence of the free
energy landscape induced by the `potential energy'.

In zero field, we obtain a simple characterization of the collinear
states selected by virtual quantum transitions. The eventual low temperature
behaviour in zero field depends on further entropic effects, and
poses an interesting open question that will be addressed
separately~\cite{US}. These results have direct
experimental relevance for the low temperature, finite field state
of high spin Kagome magnets with easy axis anisotropy, such as NGS
~\cite{kagome1,kagome2}, and perhaps other collinear Kagome magnets
as well.

{\em Model:}
Consider a layer of Kagome lattice antiferromagnet, in the limit
where the easy axis anisotropy dominates over Heisenberg exchange,
with a  `spin' $S \ge 3/2$: \be H = J\sum_{\langle ij
\rangle}\vec{S_i}\cdot\vec{S_j} - D\sum_{i}(S^z_{i})^2 -
B\sum_{i}S^z_{i}, \ee where $J > 0$ denotes the nearest neighbour
antiferromagnetic spin exchange interaction between the $S \ge 3/2$
ions, $D > 0$ is the single-ion anisotropy that picks out the
common easy axis $z$, and $B$,  the external magnetic field along
this axis, has been scaled by the magnetic moment $g\mu_B$. Due to the frustrated nature of the exchange $J$, the
anisotropy effects can begin to dominate and pick {\em collinear}
states for not-very-large $D/J$. In this collinear regime, we expect
an analysis based on the smallness of $J/D$ to give reliable
results.

With this in mind, we use $J/D$ as the small parameter in a
systematic perturbative approach that allows us to calculate the
effective low-energy Hamiltonian and the resulting low-temperature
phases. While our focus here remains the kagome case, we note
parenthetically that our methods generalize readily to closely related models of  triangular
lattice magnets in zero magnetic field; these will be discussed
separately~\cite{US}.

{\em Method:}  Our analysis proceeds by splitting the Hamiltonian as $H = H_0 + \lambda H_1$, where 
$H_0 = J\sum_{\langle ij \rangle}S^z_{i}S^z_{j} - D\sum_{i}(S^z_{i})^2 - B\sum_{i}S^z_{i}$ and $H_1 = \frac{J}{2}\sum_{\langle ij \rangle}(S^+_{i}S^-_{j} + h.c.)$. Here $S_{i}^{\pm} = S^{x}_{i} \pm iS^{y}_{i}$, and $\lambda$ is introduced as a book-keeping
device ($\lambda$ is set to one at the end of the calculation). In each case discussed below, we begin by using standard
degenerate perturbation theory~\cite{vanvleck} in $\lambda$ to
obtain the low-energy effective Hamiltonian that encodes the `slow'
dynamics induced by the $H_1$ term within the ground state manifold of $H_0$.
Terms in this expansion are naturally organized according to the
power ($n_\lambda$) of $\lambda$ they carry, and by the number
($n_b$) of {\em different} bonds on which constituents of $H_1$ act.

As it is possible to obtain a precise characterization of the pairs
$(n_\lambda, n_b)$ that contribute to leading orders in the {\em
physical} expansion parameter $J/D$, this procedure leads us
directly to the physical low-energy effective Hamiltonian at leading
orders in $J/D$. [Ref~\cite{Balentslong} employed a closely related
procedure for collinear states of the pyrochlore antiferromagnet in
a field, but without anisotropy, and our calculation is better controlled due to the presence of anisotropy.]
\begin{figure}
\includegraphics[width=\hsize]{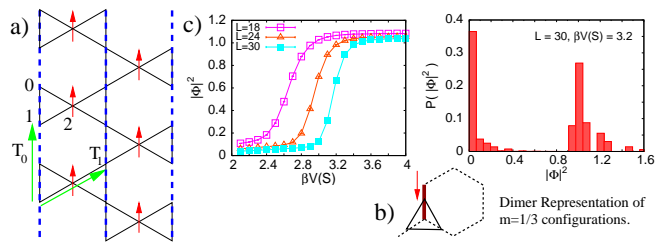}
      \caption{(color online). (a) (Sublattice) rotation symmetry breaking on the $m=1/3$ plateau: Dotted lines denote alternating arrangement of spins with average moment zero; up arrows correspond to $S_z= +S$.
c) First order transition to ordered state.}
      \label{kagome1/3}
  \end{figure}

{\it{The kagome magnet at $m=1/3$:}} We begin by considering values of field such that the ground state
manifold of $H_0$ has magnetization $m=1/3$ and is characterized by a $2:1$ constraint that
requires two spins in each triangle to be maximally polarized along
the field, and one, {\em minority}, spin in each triangle to be
maximally polarized anti-parallel to the field. While the Zeeman energy gap that drives the formation
of this magnetization plateau is largest
for $B \sim 2JS$, we will argue later that the plateau extends down to a relatively small onset field that
scales as $B_{\mathrm onset}  \sim J^3/D^2$.\begin{figure}
\includegraphics[width=\hsize]{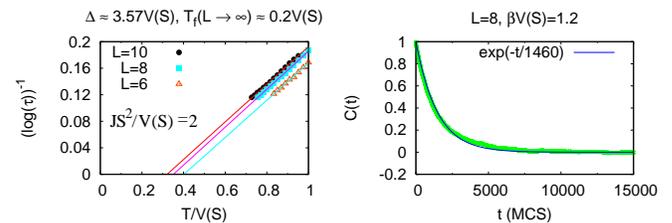}
      \caption{(color online). Data for single-spin autocorrelation time $\tau$ fit to the Vogel-Fulcher
      form $(\log(\tau))^{-1} = (T-T_f)/\Delta$; the intercept on the $x$ axis yields $T_f$, while the
      slope gives $\Delta^{-1}$.}
      \label{vogelfulcher}
  \end{figure}

A convenient way to
represent the ensemble of $m=1/3$ states~\cite{Moessner_Sondhi2} is to encode the
presence of a minority spin by placing a dimer on the corresponding
bond of the underlying honeycomb lattice (see Fig~\ref{kagome1/3} b).
The low temperature physics on the plateau is then determined by
the leading order effective Hamiltonian that acts within this dimer
subspace.

Because of the strong $2:1$ constraint, the first term
(for any $S \ge 3/2$) that breaks  degeneracy of states in the dimer
subspace is a diagonal (potential energy) term that occurs at order
$J^6/D^5$; this term arises from $n_b=6$ $n_\lambda = 6$ processes
in the Van-Vleck expansion in which the six bonds on which $H_1$
acts form a hexagonal loop. In contrast, the leading {\em
off-diagonal} term, that corresponds to changing the state of a {\em
flippable} hexagon (with alternating $\pm S$ values of $S_z$ around
the hexagon) by reversing all spins on it,  arises at order
$J^{6S-2}/D^{6S-3}$ (with $n_b=3$ $n_\lambda = 6S$).

A careful evaluation of these contributions for $S=3/2$ yields the
effective Hamiltonian: 
\be {\cal H}_{m=1/3}= \sum_{\hexagon} \left( \frac{c_1 J^6}{6D^5}|\hexagon_{1}\rangle \langle \hexagon_{1}|-\frac{c_2 J^7}{D^6}(|\hexagon_{3A}\rangle \langle
\hexagon_{3B}| + h.c.) \right)\nn \ee
where $\hexagon_m$ ($m$ = 0,1,2,3)
denote hexagonal  plaquettes with $m$ dimers and $\hexagon_{3A}$,
$\hexagon_{3B}$ represent the two flippable dimer configurations
with three dimers on a hexagonal plaquette. The values
of the coefficients are $c_1$ = $\frac{2187}{16384} \approx 0.1335$
and $c_2$ = $\frac{27}{8192} \approx 3.3 \times 10^{-3}$.  Thus, the
off-diagonal term of ${\cal H}_{m=1/3}$ is negligible
compared to the diagonal part for $S=3/2$. From the foregoing
analysis, it is also clear that its magnitude decreases with $S$,
and we have therefore not calculated the off-diagonal coefficient
for $S > 3/2$. The potential energy $V(S) \equiv c_1(S) J^6/6D^5$ has however been calculated for
general $S$, and the result is $V(S) = (2S)^6J^6/1024(2S-1)^5D^5$.

As an independent check on this result, we use the semiclassical
large-$S$ expansion procedure of Ref~\cite{Hizi_Henley} and directly
calculate the semiclassical effective Hamiltonian in this large-$S$
limit: The leading ${\mathcal
O}(J^6/D^5)$ term obtained by expanding the semiclassical result in powers of $J/D$ agrees precisely with the
large-$S$ limit of the perturbative result obtained above for
arbitrary $S$.

The low temperature physics is thus well-described by a classical dimer model with weights associated with the potential energy
term in the effective Hamiltonian.
Furthermore, this classical potential energy is minimized by any configuration with no hexagon having
precisely one dimer on it.
As there are a large number of such configurations, a more detailed analysis is needed to elucidate
the nature of the low temperature state.
To this end, we have employed a generalization~\cite{US} of the procedure of
Refs~\cite{Sandvik_Moessner,Alet_etal}  to efficiently simulate an interacting
classical dimer model with this potential energy term---our algorithm employs non-local
loop updates but preserves detailed balance in order to generate the 
correct equilibrium Gibbs distribution at temperature $T$.

As the
temperature is lowered to below $T_c \approx 0.23V(S)$ (obtained by extrapolating
the finite $L$ data in Fig~\ref{kagome1/3}), we find
that the system undergoes a transition to a state with (sublattice) rotation symmetry breaking as shown in
Fig~\ref{kagome1/3} a: In this simplest
schematic of the ordered state, one spontaneously chosen sub-lattice
of spins acquires the maximal polarization $+S$ along the field. In
order to satisfy the strong $2:1$ constraint on each triangle, the
spin moments on the other two sub-lattice
sites then alternate $+S, \; -S \dots$ in one of the two possible alternating arrangements along a stripe. [A very similar ordering
was suggested earlier for an {\em isotropic} kagome magnet in the
semiclassical limit~\cite{Hassan_Moessner}.]
This ordering can be
conveniently characterized using the sublattice order parameter $\Phi=
\sum_p m_p e^{2p\pi i/3}$, where $m_p$ denotes the sublattice
magnetization of the $p^{th}$ sublattice (Fig~\ref{kagome1/3} a).
The two peak structure in the histogram of $|\Phi|^2$ at
$T_c$ provides evidence for the first-order nature of the
transition. 

As each stripe can be in one of two possible alternating states, its internal state can be represented by an (Ising) pseudo-spin variable $\sigma$.
Are these $\sigma$ ordered in any manner,
or do they fluctuate over time? The absence of any non-zero
wavevector bragg peaks in the numerically measured static structure
factor of physical spins rules out any ${\bf q} \neq 0$ order.
Furthermore, the low temperature value of $|\Phi|^2$ is, within
error bars, exactly what one would expect if each stripe fluctuated
between its two allowed alternating states (ruling out ${\bf q} = 0$
order for the pseudospins).
Additional confirmation also comes from the statistics of  different types of hexagons with precisely two dimers on them (not shown).

This throws up a point of general interest: Although the fully equilibriated system only breaks rotation symmetry
by forming {\em disordered} stripes, this
equilibriation is achieved in our numerics because the algorithm incorporates non-local
loop updates that can flip a macroscopic number of spins in one move. In the
experimental system, the dynamics is of course purely local. 
Such local spin flips cost significant
potential energy, and the system needs to change the internal state of an entire stripe
to avoid this potential energy penalty. Systems with very similar
potential energy landscapes have been the subject of earlier studies which 
demonstrate that the time-scale for changing the internal state of a stripe diverges
rapidly with system size if the dynamics is local~\cite{Das_Kondev_Chakraborty}.
It is thus clear that the low temperature phase displays glassy freezing of the stripes.

We have also numerically studied the behaviour of the system for $T_c \lesssim T$ 
with purely local two-spin exchange dynamics satisfying detailed balance, and monitored the temperature dependence of the
single-spin autocorrelation function for a range of moderately large values for the ratio $JS^2/V(S)$.
In these simulations, the magnetic field $B$ is fixed to its nominally optimal value $B = 2JS$ which places
the system close to the center of the $m=1/3$ plateau; however, we emphasize that the ratio $JS^2/V(S)$ is kept finite as we wish to explore the higher temperature dynamics,
and configurations outside the `dimer subspace' are {\em allowed} but {\em exponentially unlikely} (as opposed to forbidden). We find that the single spin autocorrelation time $\tau$ increases very rapidly
as we lower the temperature. Indeed, $\tau (T)$ can be fit well by an activated functional form of the Vogel-Fulcher type $\tau(T) = \exp(\Delta/(T-T_f(L))$ (Fig~\ref{vogelfulcher}), thus extending further the analogy to other models of glass-formers~\cite{Das_Kondev_Chakraborty}. In our fits, the freezing temperature $T_f(L)$ drifts somewhat with linear size $L$, but its extrapolated $L \rightarrow \infty$ value
is within $10\%$ of the equilibrium $T_c$ obtained earlier, while the barrier energy scale $\Delta$ shows no $L$ dependence.

We thus conclude that while the fully equilibriated system only breaks sublattice rotation symmetry
but not lattice translation symmetry, slow glassy dynamics that sets in as the temperature is lowered
through $T_c$ forces the system into a glassy state in which the stripe pseudospins also freeze
in a {\em random} pattern, thereby breaking lattice translation symmetry in a random manner.
What would be a good experimental signature of this behaviour?
Clearly,  it is not appropriate to focus on
elastic Bragg peaks within the first Brillouin zone, as neither the fully equilibriated sublattice ordered state
nor the metastable glassy state lead to such a Bragg peak. However, we note that the sublattice order parameter that indicates the
breaking of rotation symmetry can be reconstructed by taking
suitable linear combinations of the measured spin structure factor close to
wavevectors with components $(0,0)$, $(0,2\pi/a)$, $(2\pi/a,0)$, and
$(2\pi/a,2\pi/a)$ along ${\mathrm T}_0$ and ${\mathrm T}_1$ (Fig~\ref{kagome1/3} a), and this provides a possible experimental probe of sublattice symmetry breaking, that is independent of more subtle questions
regarding the freezing of the stripe pseudospins.\begin{figure}
\includegraphics[width=0.7\hsize]{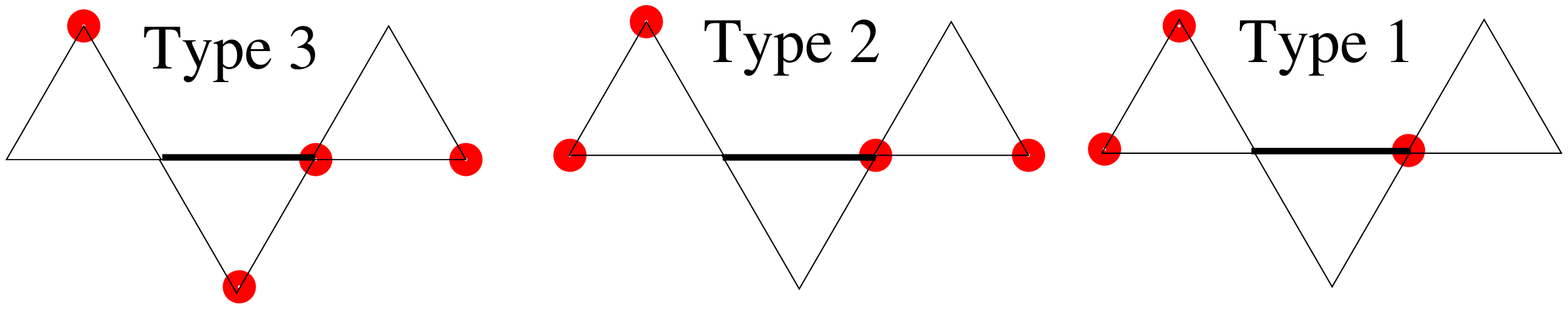}
 \caption{(color online). The three types of unfrustrated bonds; presence (absence) of red dots represents $S_z=+S$ ($S_z=-S$) }
\label{arrowproof}
 \end{figure}

{\it{The kagome magnet in zero field:}}  To understand how the onset field of the $m=1/3$ plateau scales with $J/D$ and $S$,
we need to first characterize the collinear
states selected by analogous potential energy effects in zero
field. We begin
by noting that
at large $D$ and $B=0$, the ground states of $H_0$ are obtained by requiring
all spins to have polarization $\pm S$ along the $z$ axis, and
allowing only one frustrated bond (pair of nearest neighbour
parallel spins) per triangle.

This macroscopic degeneracy is broken by the effect of real and virtual quantum transitions. A perturbative analysis in $J/D$ again allows us to derive a leading order effective Hamiltonian that acts within this degenerate subspace and encodes these effects:
\be
{\cal H}_{B=0}&=&-W\sum_{{\cal B}} (2|3 {\cal B}\rangle\langle3 {\cal B} |+ |2 {\cal B} \rangle\langle2 {\cal B} | + 0 |1  {\cal B}\rangle \langle1 {\cal B}|)\nonumber \\
&+&t \sum_{{\cal B}} (|3{\cal B}\rangle \langle 3{\cal B}^{\prime} | +
h.c.) \ee where the sum extends over unfrustrated bonds ${\cal B}$.
For $S=3/2$ we find
$W= \frac{27J^3}{64D^2}$, $t = \frac{9J^3}{32D^2}$, $|M {\cal
B}\rangle$ represent unfrustrated bonds with local environment of
type $M$  (Fig~\ref{arrowproof}), and $|3{\cal B}\rangle$
and $|3{\cal B}^{'}\rangle$ in the second (off-diagonal) term are
related to each other by a spin-exchange between the antiparallel
spins connected by the {\em flippable} ({\it i.e.} type-3) bond
${\cal B}$. More generally, for $S>3/2$, the $t$ term is negligible while $W(S)= \frac{S^3J^3}{2D^2(2S-1)^2}$.

Constraints provided by the kagome geometry allow us to
prove that the potential energy term $W$ is minimized by a class of collinear configurations
in which {\em no spin is the minority spin of
both triangles to which it belongs}~\cite{US}, and demonstrate that this class of configurations has
macroscopic entropy~\cite{US}.

More importantly for our purposes here, this zero field ensemble is stable to small magnetic
fields since the defining constraint on minority spins does not fix the average magnetization. When the field is increased further
beyond $B \sim J^3/D^2$, the ${\mathcal{O}}(BS)$ Zeeman energy gain of the $m=1/3$ ensemble (defined by the $2:1$ constraint)
will begin to dominate over the ${\mathcal{O}}(J^3S/D^2)$ potential energy gain of the zero field ensemble defined by
the constraint on minority spins, and this will trigger the onset of the $m=1/3$ plateau.
Thus,  the onset field will scale as $B_{\mathrm{onset}} \sim J^3/D^2$. Since
this onset field does not scale with $S$ (and can be quite small for even moderate values of $D/J$), we conclude that the predicted magnetization plateau state is likely to fall well within the field regime accessible to experiment even for large $S$.

{\it Discussion:} We  have thus predicted an unusual sublattice
ordered $m=1/3$ magnetization plateau state with slow {\em glassy} dynamics
at low temperature in {\em pure} $S \ge 3/2$ Kagome antiferromagnets with strong easy axis
anisotropy. We have also provided a simple characterization of the collinear states that are selected in zero external
field by an effective potential energy generated by virtual quantum fluctuations. Our results are expected to provide an excellent
starting point for understanding the very low temperature physics of
the kagome antiferromagnet NGS, although more work may be needed, particularly in the zero field case, to
understand the effects of spatial distortion and sub-dominant
in-plane anisotropies~\cite{kagome1}. We hope our
results provide impetus for experiments on other
easy axis kagome antiferromagnets, especially
members of the large family of materials~\cite{kagome1} that have  the
Ca$_3$Ga$_2$Ge$_4$O$_{14}$ crystal structure of NGS.

We would like to acknowledge useful discussions and correspondence
with L.~Balents, B.~Canals, R. Cava, D.~Dhar, R.~Moessner, T.~Senthil and
F.~Wang, computational resources of TIFR, and support from a
Hellmann Fellowship (AV), LBNL DOE-504108 (AV),
and DST SR/S2/RJN-25/2006 (KD).

\end{document}